\documentclass[10pt,journal,final,twocolumn]{IEEEtran}

\usepackage{graphics}
\usepackage{subfigure}
\usepackage{epstopdf}
\usepackage{epsfig}
\usepackage[cmex10]{amsmath}
\usepackage{amssymb}
\usepackage{times}
\usepackage{ntheorem}
\usepackage{pifont}
\usepackage{stfloats}
\usepackage{bm}
\usepackage{color}
\usepackage{makecell}
\usepackage{array}
\usepackage{multirow}

\begin{document}
%
\title{Edge Semantic Cognitive Intelligence for 6G Networks: Novel Theoretical Models, Enabling Framework, and Typical Applications}

\author{\IEEEauthorblockN{
Peihao Dong, Qihui Wu, Xiaofei Zhang, and Guoru Ding
}

\thanks{


P. Dong, Q. Wu, and X. Zhang are with the Key Laboratory of Dynamic Cognitive System of Electromagnetic Spectrum Space, Ministry of Industry and Information Technology, College of Electronic and Information Engineering, Nanjing University of Aeronautics and Astronautics, Nanjing, 211106, China (e-mail: phdong@nuaa.edu.cn; wuqihui2014@sina.com; zhangxiaofei@nuaa.edu.cn).

G. Ding is with the College of Communications Engineering, Army Engineering University, Nanjing 210007, China (e-mail: dr.guoru.ding@ieee.org).

}
\vspace{-0.3cm}
}

\IEEEtitleabstractindextext{%
\begin{abstract}
Edge intelligence is anticipated to underlay the pathway to connected intelligence for 6G networks, but the organic confluence of edge computing and artificial intelligence still needs to be carefully treated. To this end, this article discusses the concepts of edge intelligence from the semantic cognitive perspective. Two instructive theoretical models for edge semantic cognitive intelligence (ESCI) are first established. Afterwards, the ESCI framework orchestrating deep learning with semantic communication is discussed. Two representative applications are present to shed light on the prospect of ESCI in 6G networks. Some open problems are finally listed to elicit the future research directions of ESCI.
\end{abstract}

}

\maketitle

\IEEEdisplaynontitleabstractindextext

%
\IEEEpeerreviewmaketitle

\section{Introduction}

The prosperity of deep learning (DL) has endowed many areas with new vitalities. Edge computing is undoubtedly one of the beneficiaries, which incorporates deep neural networks (DNNs) to yield the unprecedentedly powerful paradigm, called edge intelligence (EI), for the sixth generation (6G) networks \cite{X. You}. Through the progressive digestion by multiple neural layers to extract the underlying features, the potential of the big data collected at the network edge can be sufficiently unleashed. Nevertheless, the real EI is not just a brute combination of DL and edge computing since this naive architecture will be overwhelmed by the explosively increasing edge data in terms of resources and latencies of communication and computing, data privacy and security, and so on \cite{X. Wang}--\cite{D. Xu}. Therefore, it is vital to elaborate the architecture in a smarter way so that the highlights of EI can be fully delivered.

\begin{figure*}[t]
\centering
\includegraphics[width=5.9in]{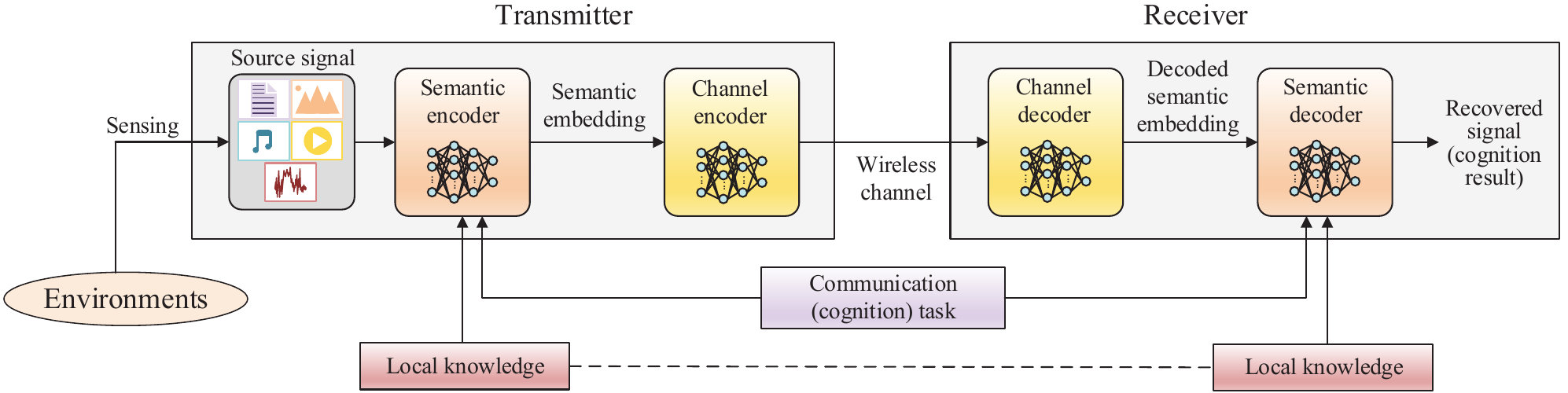}
\caption{DL-based SemCom system.}\label{DL_SemCom}
\vspace{-0.0cm}
\end{figure*}

Semantic information processing and communication may provide potential solutions to construct a more sustainable EI network. In a semantic communication (SemCom) system, the transmitter dissects the source signal and encodes the intended meaning into the message that can be digested by the receiver, and the latter decodes the received semantic message to recover the meaning, both based on the shared knowledge \cite{W. Tong}. The irrelevant information is filtered out at transmitter so that the communication resources can be saved significantly. The semantic problem of communication has been realized on the heel of the advent of Shannon information theory \cite{C. Shannon}. Carnap and Bar-Hillel spearheaded the research on semantic information based on logical probability, in distinction to the statistical perspective in Shannon information theory \cite{R. Carnap}. Along with the work in \cite{R. Carnap}, \cite{J. Bao} presented the semantic counterparts of Shannon theorems under the proposed framework for measuring semantic information. In \cite{B. Juba}, SemCom between two intelligent entities without common language or protocol was proved possible by focusing on the communication goals. In \cite{Y. Zhong}, semantic information was re-examined from the view of ecosystem. In \cite{B. Guler}, an SemCom framework impacted by an external agent was proposed and the meaning transmission problem was modeled as a Bayesian game for solution. These works provided suggestive views on semantic information theory and SemCom system design. However, they were submerged in the tidal wave of symbol transmission based on Shannon information theory since reliable symbol transmission is regarded as the primary task for communication systems in the past seven decades. Although the current communication system has almost approached Shannon capacity limit, it is still unable to satisfy the transmission of the massive data sensed by multitudinous devices in 6G networks, which calls for a more powerful communication paradigm. As the era of artificial intelligence comes, SemCom may come into play a pivotal role in 6G networks \cite{H. Xie}--\cite{Q. Hu}. In \cite{H. Xie}, a DL based SemCom (DeepSC) framework for text transmission was proposed and transfer learning was exploited to enhance the robustness in the dynamic environments. As a follow-up, the problem of speech transmission in semantic level was treated in \cite{Z. Weng} by incorporating the attention mechanism into the semantic processing DNNs. In \cite{M. Kountouris}, a goal-oriented SemCom framework inclusive of information sampling, semantic transmission, and meaning reconstruction is conceived and applied in remote actuation to show the superiority. The authors in \cite{Z. Qin} discussed semantic information theory, followed by specifying the state-of-the-art design of DL based SemCom frameworks and systems.

Fig.~\ref{DL_SemCom} illustrates a generic SemCom system empowered by DL. In more detail, the source signal, e.g., text, image, speech, video, or wireless signal collected from environments, is first processed by the semantic DNN encoder guided by the specific communication or cognition task and the common knowledge shared with the semantic decoder at the receiver. The semantic encoder can include multiple functions, such as feature extraction, meaning dissection, semantic compression and mapping, and outputs a semantic embedding, which is further processed by the DNN-based channel coding and is then conveyed via the wireless channel. At the receiver, the DNN-based channel decoder removes the impact of channel fading and physical noise on the signal. After that, the decoded semantic embedding is digested by the semantic DNN decoder to recover the desired signal, also based on the communication task and the shared knowledge with the semantic encoder. From Fig.~\ref{DL_SemCom}, the following aspects should be noted: 1) SemCom can be regarded as the cognition of environments collaborated by the transmitter and the receiver via the wireless channel, indicating that the communication task is actually the cognition task. 2) The recovered signal, also called the cognition result, is not necessarily same as the source signal according to the specific cognition goal. 3) The channel encoder and decoder can be merged into the semantic encoder and decoder to form a highly integrated system by using the more powerful DNNs. 4) The processing procedures in the SemCom system, especially in the semantic encoder and decoder, contain the highly nonlinear mapping operations and the fusion of heterogenous data, which are quite difficult to fulfill by conventional methods. So DL is anticipated to act a dominated role in designing the future SemCom systems.

Several recent works have applied SemCom to edge computing networks. In \cite{H. Xie_b}, the model training and adaption of the DeepSC were investigated to enable the semantic transmission among Internet-of-Things (IoT) devices with limited computing capability. With the similar purpose,  A semantic-aware edge network was developed in \cite{G. Shi}, where the enabling DNNs are collaboratively trained among edge severs based on federated learning. By distinguishing the difference between semantic and conventional communications and specifying the SemCom system model, three possible use cases of the SemCom in EI networks were sketched \cite{X. Luo}. In these works, edge semantic cognition intelligence (ESCI) begins to take shape but is incomplete. In this article, we will discuss theoretical models, frameworks, and applications for ESCI. Back to the radical problem, we start with clarifying the motivation of constructing EI networks from the semantic cognition perspective as follows:
\begin{itemize}[\IEEEsetlabelwidth{Z}]
\item[1)] With the help of knowledge and oriented by the cognition goal, the transmitter encodes the redundant source signal into the compact semantic embedding for wireless transmission. By doing this, both the resource occupation and the communication latency can be reduced dramatically, leading to a sustainable EI network.

\item[2)] The semantic encoding can partially or completely filter out the private information in raw data. Moreover, the receiver can successfully decode the encoded semantic information only if it has the common or at least the similar knowledge to the transmitter, which virtually forms a barrier to protect the privacy of edge data.
\end{itemize}

The rest of this article is organized as follows. Section II reviews the current semantic information theory and introduces two insightful theoretical models as bases to construct the ESCI network. The framework of the ESI network is discussed and conceived in Section III in terms of basic architectures, model training, model inference, and model adaption. Section IV provides two typical applications of ESCI in 6G networks. Section V enumerates several instructive open problems of ESCI and Section VI briefly concludes this article.

\section{Theoretical Models}

In this section, we first revisit the current semantic information theory though it is far from mature, based on which theoretical models of the semantic cognition and the semantic sampling are discussed. The established theoretical models are anticipated to provide guidances and insights for the ESCI network design.

\subsection{Semantic Information Theory}

In \cite{C. Shannon_b}, Weaver conceptually extended the technical model of communication developed by Shannon to the semantic level and effectiveness level. After that, the works on semantic information were conducted based on Shannon information theory and evolve into the current semantic information theory.

Analogical to Shannon information theory, semantic information theory starts with defining the amount of semantic information and the semantic entropy. Denote $w$ as the source signal drawn from space $\mathcal{W}$. $w$ is semantically encoded into the message, $x\in \mathcal{X}$,  and the output of the wireless channel is $y\in \mathcal{Y}$. The amount of semantic information contained in $x$ is defined as \cite{J. Bao}\footnote{Though \cite{R. Carnap} initially defined the amount of semantic information, its essence has been absorbed into the generic model of semantic information in \cite{J. Bao}.}
\begin{eqnarray}
\label{Hs_x}
H_{s}(x)=-\log_{2}p_{s}(x),
\end{eqnarray}
where $p_{s}(x)$ denotes the logical probability of $x$ and is given by
\begin{eqnarray}
\label{pl_x}
p_{s }(x)=\frac{p(\mathcal{W}_{x})}{p(\mathcal{W})}=\frac{\sum_{w\in\mathcal{W},w\vDash x}p(w)}{\sum_{w\in\mathcal{W}}p(w)}.
\end{eqnarray}
In (\ref{pl_x}), $p(\cdot)$ denotes the statistical probability and $\mathcal{W}_{x}=\{w\in\mathcal{W}|w\vDash x\}$ is the subspace in which $x$ is true. Then the semantic entropy is expressed as \cite{Q. Lan}, \cite{E. C. Strinati}
\begin{eqnarray}
\label{Hs_X}
H_{s}(X)=-\sum_{x\in \mathcal{X}} p_{s}(x)\log_{2}p_{s}(x).
\end{eqnarray}

In \cite{X. Liu}, the semantic entropy is defined sharing the similar form to (\ref{Hs_X}) from the perspective of fuzzy system as
\begin{eqnarray}
\label{Hs_zeta}
H_{s}(\zeta)=-\sum_{j=1}^{J} D_{j}(\zeta)\log_{2}D_{j}(\zeta),
\end{eqnarray}
where $D_{j}(\zeta)=\frac{\sum_{w\in\mathcal{W}_{C_j}}\mu_{\zeta}(w)}{\sum_{w\in\mathcal{W}}\mu_{\zeta}(w)}$ denotes the matching degree and $\mu_{\zeta}(w)$ denotes the membership degree of a semantic concept $\zeta$ for $w$ in class $C_j, j=1,\ldots,J$.

Based on the semantic entropy in (\ref{Hs_X}), the semantic channel capacity of a discrete memoryless channel is defined in \cite{J. Bao} as
\begin{eqnarray}
\label{Cs}
C_{s}=\sup\limits_{P(X|W)}\left\{I(X;Y)-H(W|X)+\overline{H_{s}(Y)}\right\},
\end{eqnarray}
where $P(X|W)$ is the conditional probability distribution representing a semantic coding strategy, $I(X;Y)$ denotes the mutual information between $X$ and $Y$, $H(W|X)$ denotes the ambiguity introduced by the semantic encoding, and $\overline{H_{s}(Y)}$ is the average logical information of the received messages measuring the ability of the receiver to understand the meaning of the received messages. From (\ref{Cs}), the semantic channel capacity may be either higher or lower than the Shannon channel capacity, dependent on whether the semantic interpretability of the receiver can compensate the ambiguity caused by the semantic encoder or not.

The semantic information compression was also discussed in \cite{J. Bao}. Given the entropies $H(W)$ and $H(X)$, the intuitive information loss is $L=H(W)-H(X)$. We further analyze $L$ from the semantic perspective, that is
\begin{eqnarray}
\label{Comp_Loss}
L=\underbrace{H(W)-H(\overline{Z})}_{\textrm{Intended compression}}+\underbrace{H(\overline{Z})-H(X)}_{\textrm{Lossy compression}},
\end{eqnarray}
where $\overline{Z}$ is the most concise meaning to fulfill the communication task with the help of knowledge. Thus the first term in (\ref{Comp_Loss}) denotes the intended compression to filter out the information irrelevant to the communication task and that has been interpreted by the knowledge. The second term accounts for the possible really lossy compression. Specifically, with $H(X)\geq H(\overline{Z})$, $X$ can be semantically equivalent to $\overline{Z}$ and the compression is lossless. With $H(X)<H(\overline{Z})$, it is impossible to find out an $X$ semantically equivalent to $\overline{Z}$, thus leading to a lossy compression. This insight coincides with \cite[Theorem 2]{J. Bao}.

From the sketch mentioned above, the current semantic information theory is established following the framework of Shannon information theory and the corresponding research is still in its infancy stage. Despite this, it is helpful to derive the subsequent theoretical models for the ESCI network design.

\subsection{Theoretical Model for Semantic Cognition}

The cognitive ability underlays the semantic intelligence of edge networks by understanding the meanings hidden behind the superficial information. It is essential to establish the theoretical model to evaluate the cognitive capacity \cite{Q. Wu_a}, analogous to Shannon channel coding theorem measuring channel capacity. In this subsection, we discuss this vital issue based on the semantic information theory in Section II.A.

\begin{figure}[t]
\centering
\includegraphics[width=3.0in]{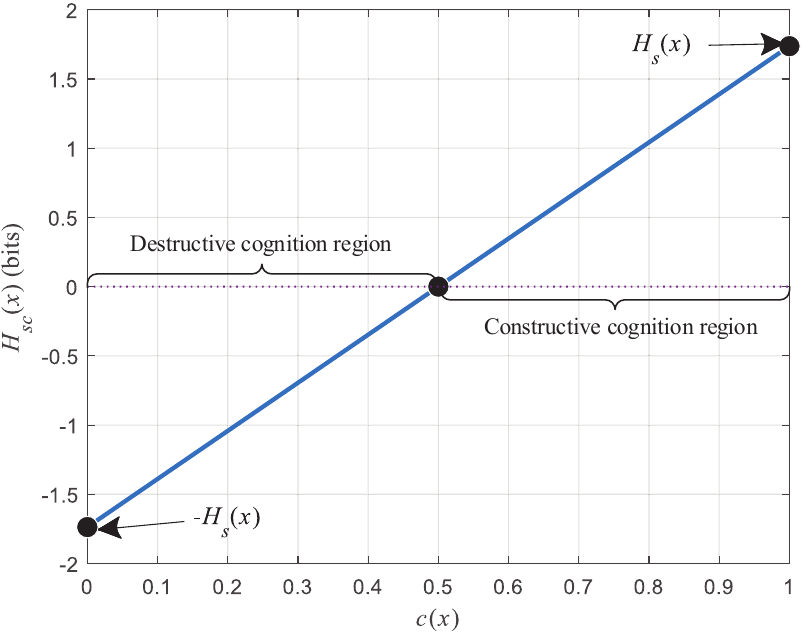}
\caption{Semantic cognitive information versus cognition accuracy.}\label{SemanCogInfo}
\vspace{-0.0cm}
\end{figure}

As pointed out in \cite{Q. Wu}, Shannon information theory is not suitable to evaluate the cognitive capacity and thus the concept of negative cognitive information was introduced. Recalling semantic message $x$ with logical probability $p_{s}(x)$, if the cognition accuracy of $x$ is $c(x)\in [0,1]$, the semantic cognitive information is defined as\footnote{The cognition correctness and the cognition accuracy should be distinguished. For the former, the cognition result is binary, i.e., correct or wrong, and is measured by the probability of correct cognition. The cognition accuracy measures how much information is correctly understood from a continuous view and is more suitable to use in semantic level.}
\setlength{\arraycolsep}{0.1em}
\begin{eqnarray}
\label{Cog_infm}
H_{sc}(x)&&=c(x)\log_{2}\frac{1}{p_{s}(x)}-(1-c(x))\log_{2}\frac{1}{p_{s}(x)}\nonumber\\
&&=c(x)H_{s}(x)-(1-c(x))H_{s}(x),
\end{eqnarray}
which measures the cognitive capacity to understand $x$ with $H_{sc}(x)>0$ and $H_{sc}(x)<0$ respectively representing constructive and destructive cognitions. Specifically, $c(x)H_{s}(x)$ and $(1-c(x))H_{s}(x)$ in (\ref{Cog_infm}) denote the semantic information acquired by correct cognition and that lost by false cognition, respectively, and the difference between them derives $H_{sc}(x)$, which monotonically increases with the cognition accuracy from $-H_{s}(x)$ to $H_{s}(x)$, as shown in Fig.~\ref{SemanCogInfo}. Consider some special cases. $c(x)=1$ leads to $H_{sc}(x)=H_{s}(x)$, indicating the cognitive system can correctly understand all the meanings of $x$. $c(x)=0.5$ leads to $H_{sc}(x)=0$, in which case the effects of correctly and wrongly understood meanings cancel each other out. $c(x)=0$ leads to $H_{sc}(x)=-H_{s}(x)$, in which case the cognitive system wrongly understands all the meanings of $x$ and thus decision making will be severely misguided.

Recalling set $\mathcal{X}$ consisted of semantic messages $x_1,\ldots,x_N$ and the semantic entropy defined in (\ref{Hs_X}), the semantic cognitive entropy is given by
\setlength{\arraycolsep}{0.1em}
\begin{eqnarray}
\label{Seman_Cog_entropy}
H_{sc}(X)&&=\sum_{i=1}^{N}c(x_i)p_{s}(x_i)\log_{2}\frac{1}{p_{s}(x_i)}\nonumber\\
&&\quad-\sum_{i=1}^{N}(1-c(x_i))p_{s}(x_i)\log_{2}\frac{1}{p_{s}(x_i)}\nonumber\\
&&\overset{(i)}{=}c(X)H_{s}(X)-(1-c(X))H_{s}(X),
\end{eqnarray}
where $(i)$ is obtained by assuming $c(x_1)=c(x_2)=\ldots=c(x_N)=c(X)$. $H_{sc}(X)$ measures the semantic cognitive capacity of a cognitive system and inherits the property of $H_{sc}(x)$ as shown in Fig.~\ref{SemanCogInfo}.

\subsection{Theoretical Model for Semantic Sampling}

In general, the data collected by edge devices contain informative redundancy. For the SemCom, the redundancy is even more under the specific communication goal and the prior knowledge. Selective data sampling can significantly lighten the burden of computing and communication as well as reduce the cost of data collection. In this subsection, the theoretical model for semantic sampling is constructed to find out the minimum number of measurements to fulfill the task in semantic level. The discussion starts with the theoretical model of sampling in data space and then extends to that of sampling in semantic space.

\emph{1) Sampling in Data Space:} Denoting $\mathbf{w}\in \mathbb{C}^{N\times 1}$ as a data vector containing the signal at each time instant or spatial location of interest in an edge network, edge devices collect only a part of $\mathbf{w}$, leading to a generic data sampling model as
\begin{eqnarray}
\label{Data_samp}
\mathbf{r}=\boldsymbol{\Theta} \mathbf{w}+\mathbf{n},
\end{eqnarray}
where the $M\times N$ binary matrix $\boldsymbol{\Theta}$ denotes the sampling operation compressing the dimension of $\mathbf{w}$ from $N$ to $M$ with $M\ll N$ and $\mathbf{n}\in \mathbb{C}^{M\times 1}$ is the additive noise at edge devices. Specifically, each row of $\boldsymbol{\Theta}$ contains exactly one non-zero entry representing the sampling time instant or spatial location and the positions of non-zero entries are non-overlapping over all rows. $\mathbf{w}$ is essentially sparse with the transform $\mathbf{w}=\boldsymbol{\Psi} \mathbf{v}$, where $\boldsymbol{\Psi}^{-1}\in \mathbb{C}^{N\times N}$ and $\mathbf{v}\in \mathbb{C}^{N\times 1}$ denote the sparse basis and the sparse form, respectively. Then (\ref{Data_samp}) can further written as
\begin{eqnarray}
\label{Data_samp_spar}
\mathbf{r}=\boldsymbol{\Theta}\boldsymbol{\Psi} \mathbf{v}+\mathbf{n}\triangleq \boldsymbol{\Phi}\mathbf{v}+\mathbf{n}.
\end{eqnarray}
By calculating the Cram\'{e}r-Rao lower bound (CRLB) on the estimation error of $\mathbf{v}$, the minimum number of measurements to reconstruct $\mathbf{w}$ under a given error can be derived.

Let us move onto a case study focusing on the spectrum mapping problem in \cite{F. Shen}, where $\mathbf{w}$, $\mathbf{v}$, and $\boldsymbol{\Psi}$ become the vector of spectral signals in all locations, the $K$-sparse vector of signal source locations, and the channel fading matrix, respectively. The CRLB on the estimation error of $\mathbf{v}$ is given by
\begin{eqnarray}
\label{CRLB_s}
\textrm{CRLB}=\frac{K}{M\bar{\gamma}\bar{\beta}},\quad\quad K\leq M\leq N,
\end{eqnarray}
where $\bar{\gamma}$ and $\bar{\beta}$ denote the average signal-to-noise ratio over all signal sources and the average large-scale fading coefficient over all channels, respectively. With a given error $\varepsilon$, the minimum number of measurements is given by
\begin{eqnarray}
\label{num_M}
M=\frac{K}{\bar{\gamma}\bar{\beta}\varepsilon}.
\end{eqnarray}
From (\ref{num_M}), if $\bar{\gamma}\bar{\beta}\varepsilon\geq 1$ holds, $M$ can take the minimum value, i.e., $M=K$, otherwise, $M$ should be larger than $K$, i.e., $M>K$.

On the other hand, the sampling operation can be regarded as a lossy compression, where $\varepsilon$ and $M$ are analogous to the distortion and the rate distortion function, respectively, defined in lossy source coding theorem in Shannon theory.

\emph{2) Sampling in Semantic Space:} Based on (\ref{Data_samp}), the semantic sampling model can be written as
\begin{eqnarray}
\label{Seman_samp}
\mathbf{r}_{s}=\boldsymbol{\Theta}_{s} \mathbf{w}_{s}+\mathbf{n}_{s},
\end{eqnarray}
where $\boldsymbol{\Theta}_{s}$, $\mathbf{w}_{s}$, and $\mathbf{n}_{s}$ respectively denote the counterparts of $\boldsymbol{\Theta}$, $\mathbf{w}$, and $\mathbf{n}$ in semantic space with the dimensions of $M_s\times N_s$, $N_s\times 1$, and $M_s\times 1$. Specifically, $\mathbf{w}_{s}$ is a feature vector. By finding out the sparse form of $\mathbf{w}_{s}$, the minimum number of measurements, $M_s$, to fulfill the task in semantic level can be calculated. Discovering the the sparse form of $\mathbf{w}_{s}$ is challenging since $\mathbf{w}_{s}$ is an abstract feature vector that is lack of interpretability. Discrete Fourier transform, discrete cosine transform, wavelet transform, or the mixed method could be tentatively exploited to solve this problem. The feature space acts as the pipeline to connect the semantic space with the data space or the physical world. Besides, based on (\ref{Seman_Cog_entropy}), the semantic cognitive entropy of $\boldsymbol{\Theta}_{s}\mathbf{w}_{s}$ can be obtained by replacing $c(X)$ with $1-\varepsilon$.


\section{ESCI Framework}

In this section, we discuss the framework of the ESCI network in terms of architectures, DNN model training, DNN model inference, and DNN model adaption by providing some new views based on the existing related works.

\subsection{Architectures}

The ESCI network mainly include two types of architectures: the device-edge-cloud (DEC)-SemCom architecture and the edge/cloud-aided device to device (D2D)-SemCom architecture.

\begin{figure}[t]
\centering
\includegraphics[width=3.0in]{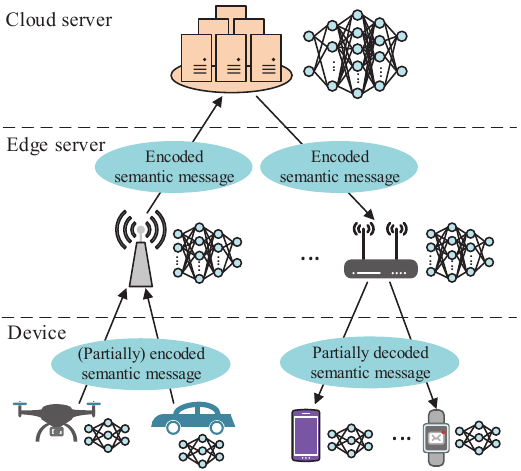}
\caption{DEC-SemCom architecture.}\label{DEC_SemCom}
\vspace{-0.0cm}
\end{figure}

\emph{1) DEC-SemCom Architecture:} Fig.~\ref{DEC_SemCom} illustrates the DEC-SemCom architecture based on the typical architecture of EI networks \cite{D. Xu}. The device first collects the source signal from the environment by using multiple sensing ways, e.g., photographing, wireless receiving, active detecting, and so on. Then the following three cases are considered. a) If the edge server is the receiver of the SemCom, the source signal is encoded by the device into the semantic message and then transmitted to the edge server. The edge server recovers the intended meaning from the received signal and makes decision accordingly. b) If the destination of the message is the cloud server, the semantic encoding can be conducted by the device and the edge server collaboratively. That is, the device encodes the source signal into a semi-finished semantic message and forwards it to the edge server to finish the remaining encoding operations. The encoded semantic message is uploaded to the cloud server for global decision making. c) If the device wants to transmit semantic information to other devices via the cloud server, the cloud server could also join in the collaborative semantic encoding. Then the cloud server transmits the encoded semantic message to another edge server and the latter partially decodes the received message. The rest of the decoding is handed over to the device to retrieve the desired meaning. For the three cases, by transmitting the concise semantic messages, the occupied resources of communication and computing can be reduced considerably. The collaborative semantic encoding and decoding between the devices and the edge servers decently invoke the computing capacities of the devices to mitigate the computing burden at the edge servers as well as to avoid exposing the private data.

\begin{figure}[t]
\centering
\includegraphics[width=3.0in]{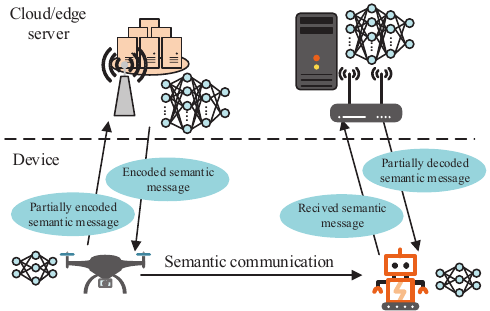}
\caption{Edge/cloud-aided D2D-SemCom architecture.}\label{EdgeCloud_D2D_SemCom}
\vspace{-0.0cm}
\end{figure}

\emph{2) Edge/Cloud-aided D2D-SemCom Architecture:} Fig.~\ref{EdgeCloud_D2D_SemCom} illustrates the edge/cloud-aided D2D-SemCom architecture developed from the semantic-aware networking model in \cite{G. Shi}. Specifically, transmitting device tentatively encodes the source signal into a semi-finished semantic message and forwards it to the edge or cloud server to finish the rest of the encoding. The encoded semantic message is then fed back to the device for semantic transmission to other device. The receiving device uploads the received semantic message to the edge or cloud server to execute most operations of the decoding. The partially decoded semantic message is then transmitted to the receiving device, in which the intended meaning is recovered by finishing the rest of the decoding. This architecture shares the similar advantages in terms of reducing overheads of communication and computing, accelerating network response, and protecting data privacy.

\begin{figure}[t]
\centering
\includegraphics[width=3.3in]{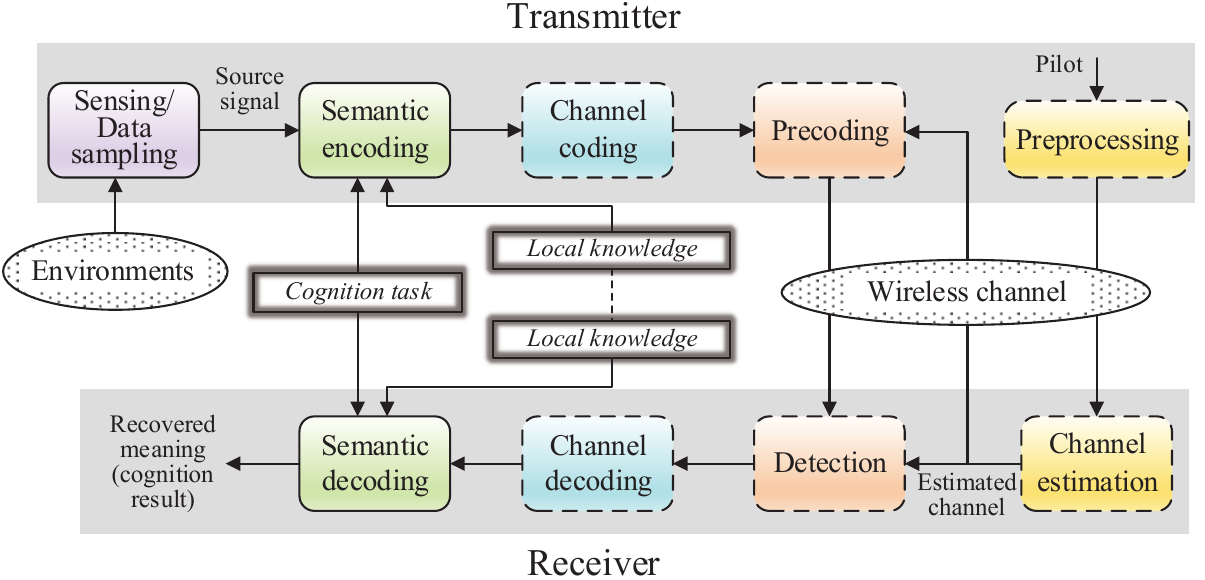}
\caption{Detailed structure of a DL-based based SemCom system.}\label{DL-SemCom-detail}
\vspace{-0.0cm}
\end{figure}

In the following, we focus on the structure of the semantic transceiver from link level, which underlays the ESCI network. The basic structure has been illustrated in Fig.~\ref{DL_SemCom}. As pointed in \cite{H. Xie_b}, dedicated channel estimation, especially DL-based approaches \cite{P. Dong}--\cite{E. Balevi}, can help equalize the impact of channel fading, which stabilizes the model training and thus improves the system performance. As a summary, Fig.~\ref{DL-SemCom-detail} illustrates the detailed structure of a DL-based based SemCom system, where each module can be driven by one or multiple properly designed DNNs. The modules with dashed lines can be skipped or merged into other modules according to the applied interface techniques and the computing capacities of devices and edge servers. It is noted that sampling in data space is adopted when collecting data from the environments while sampling in semantic space resembling semantic compression is incorporated into the semantic encoding.

\subsection{DNN Model Training}

In this subsection, we move onto model training of the enabling DNNs in the ESCI network in terms of training manner, DNN architecture, and loss function design.

\emph{1) Training Manner:} The DL-based communication system can be trained in the block-structured manner or the end-to-end manner \cite{Z. Qin_b}. For the former, the DNN for each module is separately trained via supervised learning to approximate a predesigned label, which usually needs to be elaborated by conventional methods. This training manner has been proved feasible in communication systems focusing on symbol transmission but is problematic for SemCom since the semantic encoding and decoding involve highly nonlinear mappings, leading to the difficulty in finding even a decent solution for conventional methods. Therefore, the end-to-end manner is preferred to train the semantic encoder and decoder in the current DL-based SemCom systems \cite{H. Xie}, \cite{P. Jiang}, \cite{Q. Hu}. If the modules, such as channel estimation, precoding, and detection, are considered, they can be trained separately to approximate predesigned labels or be incorporated into the end-to-end joint training framework.

\emph{2) DNN Architecture:} The core of the SemCom is extracting and processing the useful semantic information from the source signal. Some typical works on natural language processing tasks have applied the recurrent neural network and the convolutional neural network (CNN) to understand sentence meanings \cite{A. Graves}, \cite{N. Kalchbrenner}, but perform unsatisfactorily in face of long sentences. In \cite{H. Xie}, Transformer with the attention mechanism \cite{A. Vaswani} was exploited to design the semantic encoder and decoder for text transmission and achieved the superior performance. The vision Transformer based semantic encoder and decoder were developed in \cite{Q. Hu} for image identification over the air with input masking. However, the minimum number of unmasked input entries required for image reconstruction or classification is not investigated, which can be derived based on the theoretical model for sampling in Section II.C. As mentioned above, some modules can be combined and driven by one DNN. If combining the semantic encoding/decoding with the channel encoding/decoding \cite{P. Jiang}, the end-to-end error can be reduced remarkably compared to the separate design. With the help of the attention mechanism, the integrated encoder and the integrated decoder were designed for speech transmission in \cite{Z. Weng} based on the residual network with squeeze-and-excitation. Merging DNNs of multiple modules could reduce the computational complexity but undoubtedly deteriorates the interpretability of the DNN, which needs further investigation, especially when using multiplexing techniques. The DNN architectures of precoding, detection, and channel estimation have been elaborated for DL-based communication systems in technical level \cite{A. M. Elbir}--\cite{P. Dong_b}, which can be exploited to design the counterpart modules in Fig.~\ref{DL-SemCom-detail} with proper modification. For EI networks, the available computing resources should be taken into consideration for DNN architecture design. Specifically, the BranchyNet was proposed in \cite{S. Teerapittayanon}by adding several lite branches with fewer neural layers at the specific points of the original DNN, based on which the devices and edge servers can choose the proper DNN architecture according to the real-time available computing resources.

\emph{3) Loss Function Design:} The loss function orients the DNN training and thus is vital for the performance. For DL-based SemCom systems, the loss function is designed as per the specific communication task. Considering sentence transmission, the loss function designed in \cite{H. Xie} consists of two parts for respective optimization goals. The cross-entropy (CE) part minimizes the difference between the original and recovered sentences while the mutual information part aims to maximize the transmission rate with a weight balancing the two parts. The sentence semantic transmission in \cite{P. Jiang} divides the training process into three stages with loss functions of CE, mean-squared error (MSE), and CE, respectively. For image classification via the wireless channel, the loss function designed in \cite{Q. Hu} includes three components to collaboratively optimize the encoder and decoder as well as the basis parameters. In \cite{Z. Weng}, MSE was applied as the loss function, which is proved effective for speech signal reconstruction. If the communication task is to cognize the propagation environments, the semantic cognitive entropy in (\ref{Seman_Cog_entropy}) can be used to design the loss function. The loss function is usually related to the performance metric of the SemCom system. Thus the performance metric could provide insights on loss function design, e.g., the bilingual evaluation understudy score \cite{K. Papineni}, perceptual loss in VGG feature space \cite{J. Johnson}, perceptual evaluation of speech distortion \cite{A. W. Rix}, and so on.

\subsection{DNN Model Inference}

In the following, we discuss the issues of DNN model inference when the offline trained DNNs are used to fulfill the tasks in the practical ESCI network.

\emph{1) DNN Model Compression:} To accommodate the limited computing resource and the latency requirement in the inference stage, the offline trained DNNs usually need to be compressed. Network pruning and weight quantization are two widely used compression approaches \cite{S. Han}--\cite{J. Wu}. Network pruning aims to cut off the dispensable connections between neurons while maintaining a satisfactory accuracy. In \cite{S. Han}, the weights less than the threshold were regarded as trivial contributors to the performance and are thus removed from the DNN, after which the remaining weights are fine-tuned to compensate the performance loss. In \cite{Y. Gong}, the weights of a CNN were clustered by K-means and are then quantized, which reduces the DNN size by up to 24$\times$ while the accuracy loss is negligible. Also using K-means method, the clustered weights were quantized under the criterion of minimizing the estimated error of each layer \cite{J. Wu}. In the recent work on SemCom \cite{H. Xie_b}, both network pruning and weight quantization were applied to enable the device with limited computing capacity to execute the DNN-driven semantic encoding. In addition, by adopting the DNNs such as BranchyNet \cite{S. Teerapittayanon}, the device and edge/cloud server can select a proper branch from among the alternatives to accommodate the real-time available computing resources, which can be deemed as a dynamic way for model compression.

\emph{2) DNN Model Deployment:} Another key of model inference is to properly deploy the DNNs in the ESCI network, which, however, is not straightforward since it is dependent on the specific communication requirement and the computing capacity. If the device (edge server) and the edge server (cloud server) are the transmitter and the receiver, respectively, the DNNs can be directly deployed in the regular way. If the semantic encoding or decoding is conducted collaboratively in the network, DNN model partition should be considered. That is, the trained DNN for semantic encoding/decoding is split into two or more parts, which are then deployed at different locations of the network, as shown in Fig.~\ref{DEC_SemCom} and Fig.~\ref{EdgeCloud_D2D_SemCom}. The criterion to determine the partition position in the DNN is mainly dependent on the resources and latency requirements of computing and communication. In \cite{Y. Kang}, a regression-based approach was proposed to evaluate the runtime of each layer, based on which the optimal partition position is obtained under the latency or energy constraint. With the requirement of accuracy, the integer linear programming method was leveraged in \cite{H. Li} to find out the partition position achieving the minimal inference latency.



\subsection{DNN Model Adaption}

In face of dynamic environments and tasks in the EI network, DNN model adaption is necessary to guarantee robustness. We will discuss several enabling methods for model adaption, i.e., transfer learning, cognitive learning, and federated learning.

\emph{1) Transfer Learning:} The new environment or task shares the underlying similarity to the one that has been seen by the DNN. Thus the knowledge learnt by the DNN can be transferred to the new case to enable a fast model adaption \cite{C. Tan}. DNN weight transfer is a widely applied approach. Specifically, the shallow layers of the pre-trained DNN are retained as a versatile feature extractor while the remaining part is fine-tuned based on data collected in the target domain. By adopting transfer learning, the number of weights need to be retrained online is usually minor, leading to a considerable reduction in the computing overhead for the ESCI network to adapt to new situations. Besides, for the collaborative semantic encoding in Fig.~\ref{DEC_SemCom} and Fig.~\ref{EdgeCloud_D2D_SemCom}, the shallow layers are executed at the device and could be retained in the online adaption stage, which further mitigates the computing burden of the device.

\emph{2) Cognitive Learning:} Inspired by the powerful brain cognitive mechanism, a unified cognitive learning framework was proposed in \cite{Q. Wu_b}, which can flexibly choose the appropriate learning algorithm to adapt to the dynamic wireless environments and tasks. By incorporating the self-learning and the online adaption of cognitive learning into the SemCom system, the performance of the ESCI network could be further boosted.

\emph{3) Federated Learning:} Under the rapidly changing wireless environment, the network should collect sufficient data for fine-tuning and finish the adaption much faster than the variation of wireless statistics. Federated learning provides a potential way to tackle this challenge by integrating the collected training data and computing resources of multiple devices or edge servers for collaborative online adaption \cite{W. Y. B. Lim}. In \cite{G. Shi}, federated learning was exploited to develop a semantic-aware network, where the locally trained DNN weights of several edge servers are aggregated in a coordinator to accelerate the online training and protect the data privacy. Moreover, federate learning has been combined with meta-learning to further speed up the model adaption \cite{F. Chen}, \cite{A. Fallah} and could be considered to design the DNN training strategy of the ESCI network. Model or gradient compression is pivotal to enable transmissions of DNN weights or loss gradients for federate learning based model adaption. In addition to the model compression methods mentioned above, some typical gradient compression methods have been proposed in \cite{Y. Lin}--\cite{H. Tang} by using momentum correction and factor masking, local gradient clipping, gradient update sparsification, gradient quantization, and so on.


\section{Applications of ESCI}

In this section, we elaborate two representative applications of ESCI, i.e., semantic cognition of spectrum and semantic vehicle-to-everything (V2X) communication, derived from the basic architectures introduced in Section III.A.

\subsection{Semantic Cognition of Spectrum}

\begin{figure}[t]
\centering
\includegraphics[width=3.0in]{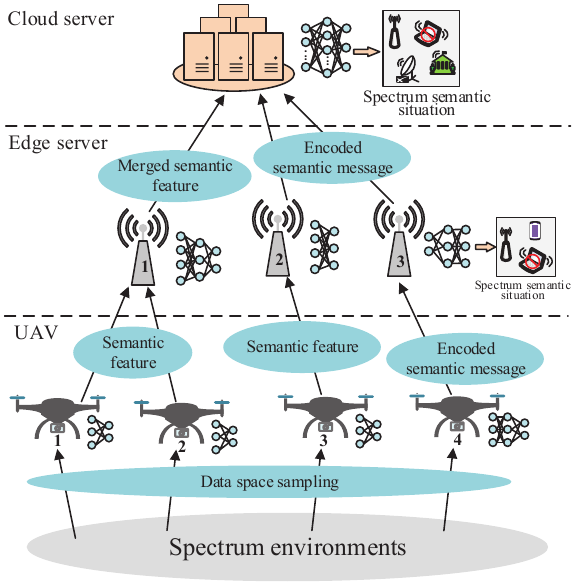}
\caption{UAV swarm assisted spectrum semantic cognition network.}\label{App1}
\vspace{-0.0cm}
\end{figure}

Fig.~\ref{App1} illustrates a spectrum semantic cognition network assisted by the unmanned aerial vehicle (UAV) swarm. In distinction to the conventional spectrum situation presenting just shallow information of signal strength, the spectrum semantic situation reveals more compact while insightful meanings that are more convenient to transmit and exploit for the subsequent decision making. In Fig.~\ref{App1}, the UAVs collect data from the spectrum environment based on the data space sampling model in (\ref{Data_samp}). UAVs 1 and 2 use the equipped lite DNNs to preprocess the collected data and then transmit the shallow semantic features to edge server 1. Edge server 1 merges the received features into a more concise semantic feature and uploads the output to the cloud server to participate in constructing the global semantic situation of the spectrum. UAV 3 partially encodes the collected data into the semantic feature, which is then handed over to edge server 2 for the remaining semantic encoding operations, including semantic sampling and mapping. UAV 4 transforms the collected data into the fully encoded semantic message and then transmits it to edge server 3. The latter decodes the received semantic message and constructs a local semantic situation of the spectrum. The encoded semantic messages at edge servers 2 and 3 will also be forwarded to the cloud server to construct the global semantic situation of the spectrum collaborating with other uploaded semantic information. From Fig.~\ref{App1}, for example, the spectrum semantic situation is able to tell us the location of a pseudo base station or a covert enemy headquarters through the hierarchical and collaborative processing among UAVs, edge servers and cloud server.

\subsection{Semantic V2X Communication}

\begin{figure}[t]
\centering
\includegraphics[width=3.0in]{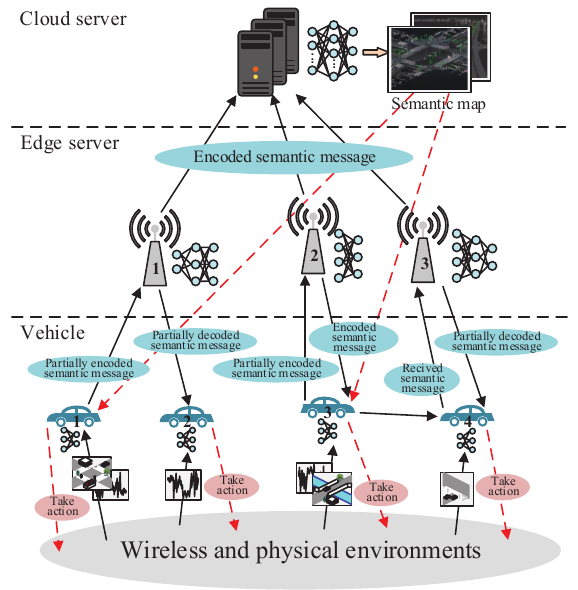}
\caption{SemCom-based V2X network.}\label{App2}
\vspace{-0.0cm}
\end{figure}

Fig.~\ref{App2} illustrates a SemCom-based V2X network, where vehicles, edge servers, and cloud server interact semantic information to facilitate human and automatic driving. The vehicle 1 senses the surrounding wireless and physical environments and intends to share the useful information with vehicle 2. Considering the limited communication and computing resources, the information is partially encoded into the compact semantic message and is then forwarded to edge server 1 for further processing. The output of the DNN at edge server 1 can be regarded as a partially decoded semantic message, based on which, vehicle 2 can recover the desired semantic information readily by merging some information collected by itself and then takes action. Similar to vehicles 1 and 2, vehicles 3 and 4 also want to interact the information about the wireless and physical environments while the communication between them follows the architecture shown in Fig.~\ref{EdgeCloud_D2D_SemCom}. In addition, all the encoded semantic messages at edge servers are uploaded to the cloud server to generate the overall semantic map, which in turn instructs vehicles to take action.

ESCI can be used in a wide range of applications, e.g., industrial IoT, virtual reality, and smart medical service, and we just discuss two typical applications therein. The detailed working processes of these promising applications desire in-depth investigation.

\section{Future Research Topics}

ESCI is still in its infancy and here we enumerate the key open problems and future research directions:
\begin{itemize}[\IEEEsetlabelwidth{Z}]
\item[1)] \textbf{Theoretical Model:} For ESCI, theoretical models are pivotal in exploring mechanisms, designing algorithms, and evaluating performance, but the corresponding research is in the initial stage. The future investigation can focus on the new framework for semantic information theory, the general form of semantic cognitive entropy, the rigorous modeling of semantic sampling, and so on.

\item[2)] \textbf{Communication Efficiency and Computing Resource Tradeoff:} Semantic communication reduces transmission overheads at the cost of computing resources dedicated to the more complicated DNNs. As the computational complexity of DL has far exceeded the Moore's Law, it is vital to explore the fundamental tradeoff between communication efficiency and computing resource to guarantee the sustainability of ESCI.

\item[3)] \textbf{Transceiver Architecture:} The DL-based semantic transceiver includes multiple modules. It needs to discover the mechanism of module integration by considering the applied interface techniques and the computing capacities of devices and edge servers.

\item[4)] \textbf{Knowledge Update:} Knowledge shared by the transceiver is one of key enablers underlaying ESCI. It is important to distinguish different types of knowledge, e.g., long-term knowledge and short-term knowledge, and to update them in the proper manner and frequency.

\end{itemize}

\section{Conclusion}

In this article, we discuss and conceive ESCI in the aspects of theoretical models, framework, and applications. Although these views are far from mature, they shed light on the attractive potential of ESCI and we hope they can spur escalating research interest on ESCI for 6G networks.



\ifCLASSOPTIONcaptionsoff
  \newpage
\fi

\end{document}